\documentclass[galaxies,letter,accept,moreauthors,pdftex,10pt,a4paper]{mdpi}
\firstpage{1}
\articlenumber{39}
\doinum{10.3390/galaxies5030039}
\pubvolume{5}
\pubyear{2017}
\copyrightyear{2017}
\externaleditor{Academic Editors: Duncan A. Forbes and Ericson D. Lopez}
\history{Received: 29 June 2017; Accepted: 9 August 2017; Published: 15 August 2017}

\Title{Deep MOS Spectroscopy of NGC\,1316 Globular~Clusters}

\Author{Leandro A. Sesto $^{1,2,*}$, Favio R. Faifer $^{1,2}$, Juan C. Forte $^{3,4}$ and Anal\'{\i}a V. Smith Castelli $^{2}$}

\AuthorNames{Leandro Sesto, Favio R. Faifer, Juan C. Forte and Anal\'{\i}a V. Smith Castelli}

\address{%

$^{1}$ \quad Facultad de Ciencias Astron\'{o}micas y Geof\'{\i}sicas,  Universidad Nacional de La Plata , Paseo del Bosque s/n, La Plata B1900FWA, Argentina; favio@fcaglp.unlp.edu.ar\\

$^{2}$ \quad Instituto de Astrof\'{\i}sica de La Plata (IALP; CCT La Plata, CONICET-UNLP), Paseo del bosque s/n, \mbox{La Plata B1900FWA}, Argentina; asmith@fcaglp.unlp.edu.ar\\

$^{3}$ \quad The National Scientific and Technical Research Council (CONICET), CABA C1425FQB, Argentina; forte@fcaglp.unlp.edu.ar\\

$^{4}$ \quad Planetario de la Ciudad de Buenos Aires, CABA C1425FGC, Argentina\\}

\corres{Correspondence: sesto@fcaglp.unlp.edu.ar; Tel.: +54-221-483-7324}

\abstract{The giant elliptical galaxy NGC\,1316  is the brightest galaxy in the Fornax cluster, and displays a number of morphological features that might be interpreted as an intermediate age merger remanent ($\sim$3 Gyr). Based on the idea that globular clusters systems (GCS) constitute genuine tracers of the formation and evolution of their host galaxies, we conducted a spectroscopic study of approximately 40 globular clusters (GCs) candidates associated with this interesting galaxy. We~determined ages, metallicities, and $\alpha$-element abundances for each GC present in the sample, through the measurement of different Lick indices and their subsequent comparison with simple stellar populations models~(SSPs).}


\begin{document}



\section{Introduction}

The giant elliptical galaxy and strong radio source NGC\,1316 displays a number of morphological features that might be interpreted as a merger remnant of approximately 3 Gyr (\citep{Goudfrooij2001a}). Among them, we can emphasise shells, ripples, and an unusual pattern of dust, formed by large filaments and dark structures. This galaxy is located at a distance of 20.8 Mpc (\citep{Cantiello2013}), and belongs to Fornax, one of the closest and most studied galaxy clusters of the southern hemisphere.

\begin{figure}[H]
\center
	\includegraphics[width=0.7\columnwidth]{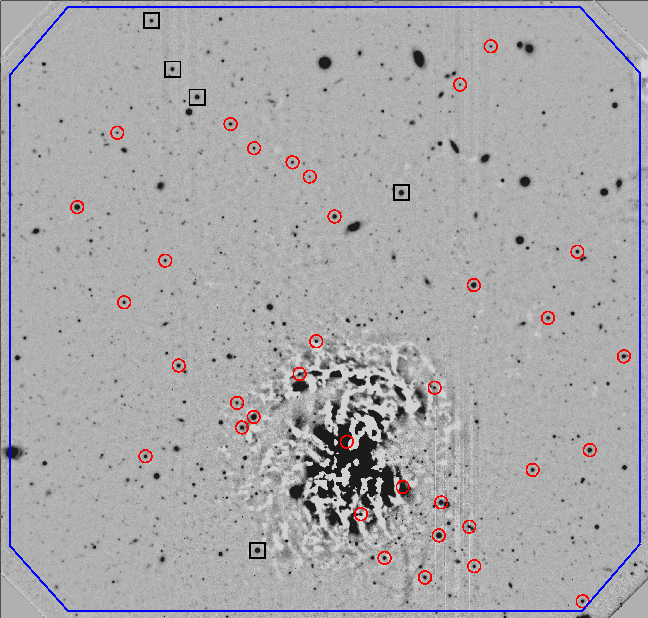}
    \caption{Image used as pre-image for spectroscopy in which the halo of NGC\,1316 has been subtracted. Confirmed GCs are shown with red circles and field stars with black squares (see Section \ref{sec3}). The North is up and East to the left. The field of view (FOV) is 5.5'$\times$5.5'.}
    \label{fig1}
\end{figure}

In a previous photometric work, we detected the presence of different globular clusters (GCs) sub-populations likely associated with different merger events (\citep{Sesto2016}). In this context, we conducted a spectroscopic study of 40 globular clusters candidates belonging to NGC\,1316 using the multi-object mode of the Gemini Multi-Object Spectrograph (GMOS), mounted on the Gemini South telescope (Figure \ref{fig1}). As a result of good quality data and a detailed reduction, we have obtained spectra with excellent signal-to-noise ratio (S/N) (some of them with S/N \textgreater 50). This allowed us to determine radial velocities, ages, metallicities, and $\alpha$-element abundances for each GC present in the sample (Sesto et al., in preparation).

\section{Observations and Data Reduction}

The data were obtained between August 2013 and January 2014 as part of the Gemini program GS-2013B-Q-24 (PI: Leandro Sesto). The MOS mask consisted of 40 slits of 1 arcsec width and 4--6~arcsec length. We used the B600-G5303 grating centered at 5000 and 5100 \AA~(to cover the CCD chip gaps), with $2\times2$ binning, and we considered exposure times of $16\times1800$ s, yielding 8 hours of on-source integration time. The obtained MOS spectra typically covers the range 3500--6500 \AA, with a dispersion of 0.90 \AA/pix and a spectral resolution of approximately 4.7 \AA. The images were processed using the GEMINI-GMOS routines in IRAF. This process was carried out in different stages, which~included corrections by bias and flat field, calibration in wavelength, and the extraction and subsequent combination of individual spectra. Finally, spectroscopic standard star observations were used to transform our instrumental spectra into flux-calibrated spectra.
\section{Confirmation of GCs}
\label{sec3}
The radial velocities (RVs) of the sample objects were determined through cross-correlation with different synthetic models of simple stellar populations (SSPs) using the method of \citep{Tonry1979}, incorporated in the FXCOR task within IRAF. As template spectra we used stellar population synthesis models obtained from MILES libraries (\citep{Vazdekis2010}). A total of 19 SSP models were considered, covering a wide range of ages (2.5, 5, and 12.6 Gyr) and metallicities ([Z/H] = $-2.32; -1.71; -1.31; -0.71; -0.4 ; 0; 0.4$ dex), with a unimodal initial mass function (IMF) with a slope value of 1.3.

Thirty-five genuine globular clusters were confirmed, which present radial velocities close to 1760~km/s, adopted as the systemic velocity of NGC\,1316 (\citep{Longhetti1998}). Only five objects presented in the sample were field stars with heliocentric radial velocities lower than 60 km/s.

\section{Lick/IDS~Indices}

In order to determine the age, metallicity, and $\alpha$-element abundances of the GCs, the $\chi^2$ minimization method of \citep{Proctor2002} and \citep{Proctor2004} was used. This technique simultaneously compares the different observed Lick/IDS indices with those obtained from simple stellar population models, selecting the combination that minimizes the residuals through a $\chi^2$ fitting. In this particular case, we used the SSP models of \citep{Thomas2003,Thomas2004}, which have a spectral coverage between 4000 and 6500 \AA, ages from 1 to 15 Gyr, and metallicities of [Z/H] = $-$2.25 to 0.67 dex. One of the most outstanding features of these models is the fact that they include the effects produced by the abundance relations of $\alpha$-element. These models consider [$\alpha$/Fe] = 0.0, 0.3, 0.5 dex.

To estimate the integrated properties of each GC, we used those spectral indexes that presented the smallest errors. These were selected from the group conformed by H$\delta_A$, H$\delta_F$, H$\gamma_A$, G4300, Fe4383, H$\beta$, Fe5015, Mgb, Fe5270, Fe5335 and Fe5406, since they provide acceptable results for the study of extragalactic GCs (\citep{Norris2006}).

\section{Ages, Metallicities, and $\alpha$-element Abundances}

\begin{figure}[h]
\center
	\includegraphics[width=0.7\columnwidth,angle=270]{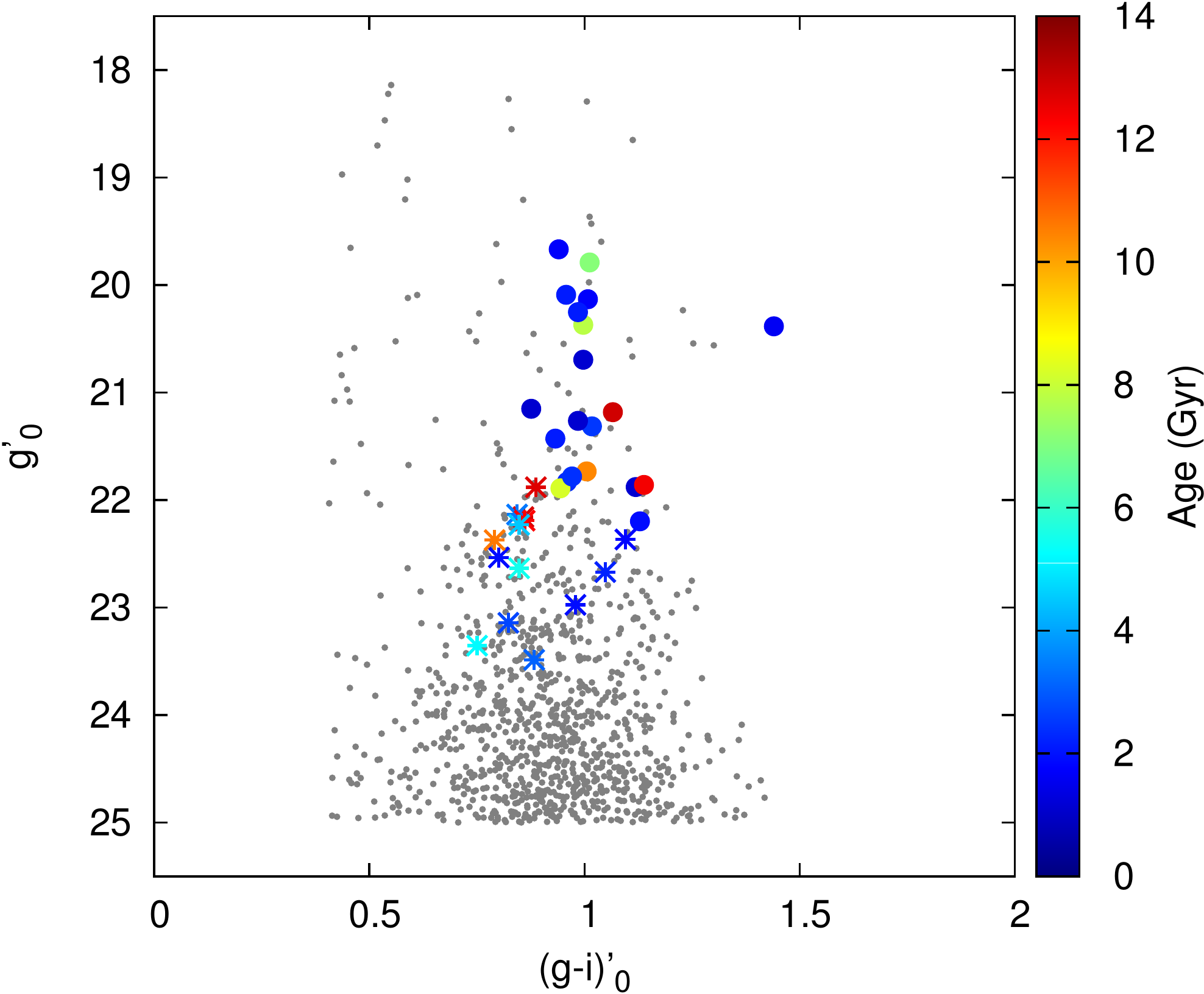}
    \caption{Color-magnitude diagram of the photometric GCs candidates (\citep{Sesto2016}). The grey dots correspond to the globular clusters (GCs) candidates brighter than $g'_0 =$ 25\,mag. The colors indicate the different ages of the of the 35 spectroscopically confirmed GCs. We distinguish between GCs with spectra with signal-to-noise ratio (S/N) \textless 25 (stars) and S/N  \textgreater 25 (circles).}
    \label{fig2}
\end{figure}

Figure \ref{fig2} shows the color-magnitude diagram of the photometric Gemini-$g'r'i'$ data (corrected for interstellar extinction) presented in \citep{Sesto2016}, where colors indicate the different ages of the GCs with spectroscopic information. The figure shows the presence of an important group of young GCs with ages close to 2 Gyr.

The ages of some GCs associated with NGC\,1316 were previously established by \citep{Goudfrooij2001a}. These~authors obtained spectra for three GCs with S/N good enough to determine ages and metallicities. Only one of them was present in our spectroscopic sample. A significant discrepancy was observed with respect to the age of this object, as these authors estimated in $\sim$3$\pm$0.5 Gyr, whereas in this work we measured an age of 7.1$\pm$0.4 Gyr. It is important to note that the spectrum used in this work presents a considerably higher S/N value, which allowed us to measure many more spectral indicators than those used by \citep{Goudfrooij2001a}. This result is particularly interesting, since it could be indicating that this GC was not formed in the last merger event experienced by NGC\,1316.

\begin{figure}[H]
\center
	\includegraphics[width=0.7\columnwidth,angle=270]{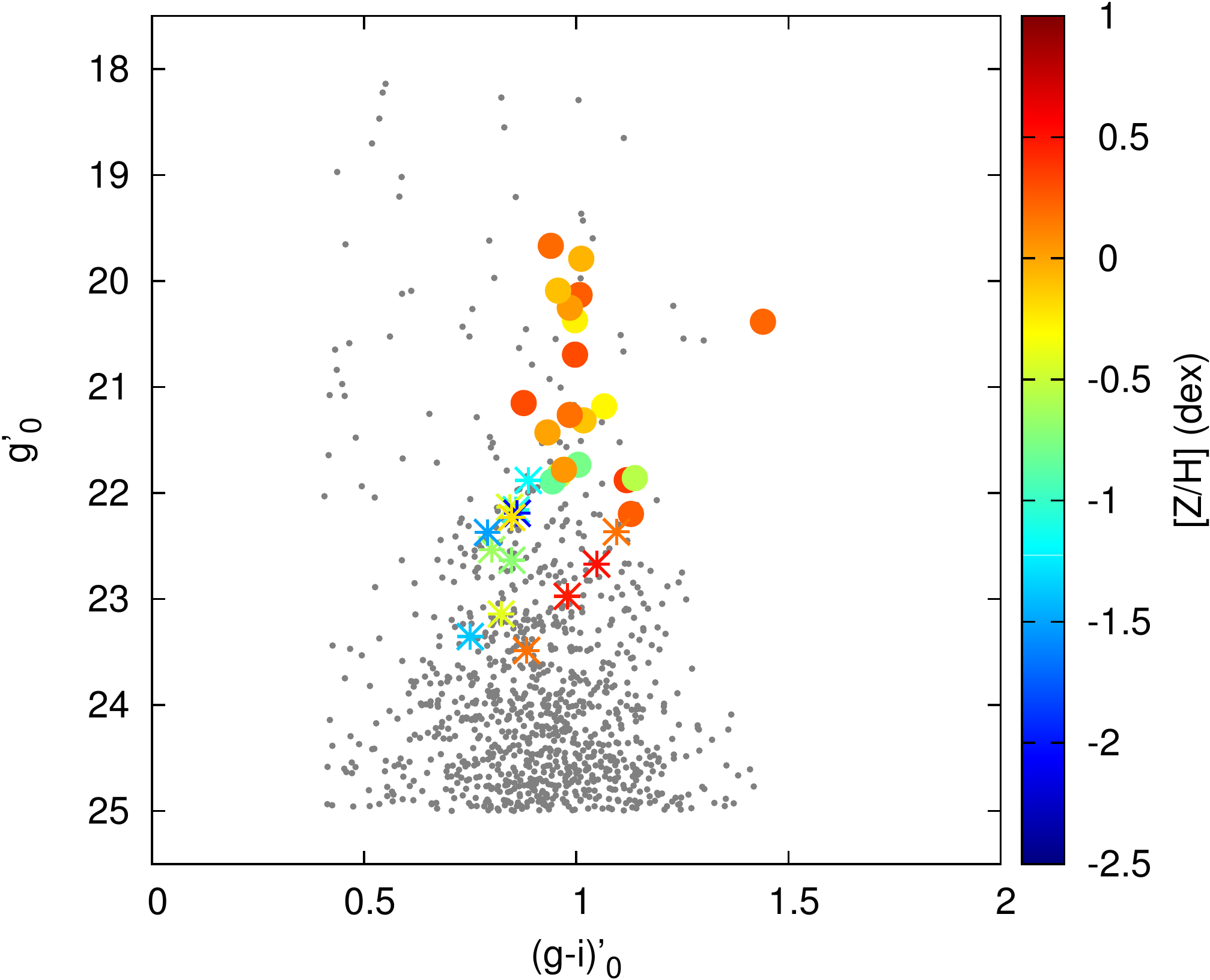}
    \caption{Color-magnitude diagram of the photometric GCs candidates (\citep{Sesto2016}). The grey dots correspond to the GCs candidates brighter than $g'_0 =$ 25\,mag. The colors indicate the different metallicities of the 35 spectroscopically confirmed GCs. We distinguish between GCs with spectra with S/N \textless 25 (stars) and S/N  \textgreater 25 (circles).}
    \label{fig3}
\end{figure}

Figure \ref{fig3} shows the color-magnitude diagram of the photometric and spectroscopic data, where colors indicate the different metallicities ([Z/H] measured in dex) of the 35 spectroscopically confirmed GCs. In the particular case of objects with S/N \textgreater 25, the sample is dominated by objects with relatively high metallicities; i.e., $-0.5 \textless [Z/H] \textless 0.5$ dex.

\section{Conclusions}

The spectroscopic results have confirmed the presence of multiple GC populations associated with NGC\,1316, among which stands out the presence of a population of young GCs with an average age of 1.7 Gyr and metallicities between $-0.5 \textless [Z/H] \textless 0.5$ dex. These results will be analyzed in a future work with the aim of describing the different episodes of star formation, and thus at obtaining a more complete picture about the evolutionary history of the galaxy.

\vspace{12pt}

\authorcontributions{{All authors contributed equally to this work.}}
\conflictsofinterest{The authors declare no conflict of interest.}

\reftitle{References}


\end{document}